\documentclass[aps,superscriptaddress,twocolumn,twoside,floatfix,pra,longbibliography,a4paper,nofootinbib,accepted=2022-03-10]{quantumarticle}
\pdfoutput=1
\usepackage{epsfig}
\usepackage{amsfonts}
\usepackage{amsmath}
\usepackage{amssymb}
\usepackage{amsthm}
\usepackage{color}
\usepackage{multirow}
\usepackage[normalem]{ulem}
\newcommand{\stkout}[1]{\ifmmode\text{\sout{\ensuremath{#1}}}\else\sout{#1}\fi}
\usepackage{latexsym}
\usepackage{mathrsfs}
\usepackage{natbib}
\usepackage{verbatim}
\usepackage[T1]{fontenc}
\usepackage{float}
\usepackage{subfigure}
\usepackage[table,xcdraw]{xcolor}
\usepackage{lipsum}

\usepackage{pifont}

\usepackage{graphicx}
\usepackage{xcolor}

\usepackage[colorlinks=true,linkcolor=blue,citecolor=magenta,urlcolor=blue]{hyperref}

\definecolor{darkorange}{RGB}{255,100,0}

\DeclareMathOperator{\Tr}{tr}
\newcommand{\A}{\text{A}}
\newcommand{\B}{\text{B}}
\newcommand{\gA}{\gamma_\text{A}}
\newcommand{\gB}{\gamma_\text{B}}
\newcommand{\TA}{T_\text{A}}
\newcommand{\TB}{T_\text{B}}

\newcommand{\ket}[1]{|#1\rangle}

\newcommand{\bracket}[3]{\langle#1|#2|#3\rangle}
\newcommand{\ketbra}[2]{|#1\rangle\langle#2|}

\newcommand{\armin}[1]{\textit{\small\textcolor{red}{Armin: #1}}}

\begin{document}


\title{Operational nonclassicality in minimal autonomous thermal machines}


\author{Jonatan Bohr Brask}
\affiliation{Department of Physics, Technical University of Denmark, 2800 Kongens Lyngby, Denmark}

\author{Fabien Clivaz}
\affiliation{Institut f{\"u}r Theoretische Physik und IQST, Universit{\"a}t Ulm, Albert-Einstein-Allee 11, D-89069 Ulm, Germany}
\affiliation{Institute for Quantum Optics and Quantum Information - IQOQI Vienna, Austrian Academy of Sciences, Boltzmanngasse 3, 1090 Vienna, Austria}

\author{G\'eraldine Haack}
\affiliation{D\'epartement de Physique Appliq\'ee, Universit\'e de Gen\`eve, 1211 Geneva, Switzerland}

\author{Armin Tavakoli}
\email{armin.tavakoli@oeaw.ac.at}
\affiliation{Institute for Quantum Optics and Quantum Information - IQOQI Vienna, Austrian Academy of Sciences, Boltzmanngasse 3, 1090 Vienna, Austria}
\affiliation{Atominstitut,  Technische  Universit{\"a}t  Wien, Stadionallee 2, 1020  Vienna,  Austria}

\begin{abstract}
Thermal machines exploit interactions with multiple heat baths to perform useful tasks, such as work production and refrigeration. In the quantum regime, tasks with no classical counterpart become possible. Here, we consider the minimal setting for quantum thermal machines, namely two-qubit autonomous thermal machines that use only incoherent interactions with their environment, and investigate the fundamental resources needed to generate entanglement. Our investigation is systematic, covering different types of interactions, bosonic and fermionic environments, and different resources that can be supplied to the machine. We adopt an operational perspective in which we assess the nonclassicality of the generated entanglement through its ability to perform useful tasks such as Einstein-Podolsky-Rosen steering, quantum teleportation and Bell nonlocality.  We provide both  constructive examples of nonclassical effects and general no-go results that demarcate the fundamental limits in autonomous entanglement generation. Our results open up a path toward understanding nonclassical phenomena in thermal processes. 
\end{abstract}

\maketitle


\section{Introduction}

Classical thermal machines, such as the steam engines which emerged with the industrial age, produce work, heating, or cooling, by exploiting heat currents between environments at different temperatures. The study of fundamental performance limits for such machines was integral to the development of the theory of thermodynamics. Similarly, the study of quantum thermal machines plays a key part in ongoing research aiming to advance understanding of thermodynamics in the quantum regime \cite{Binder2018, Bhattacharjee2020}. Such machines incorporate quantum effects in their operation but function according to the same basic modus operandi. It is hence a compelling question whether there is a fundamental sense in which quantum thermal machines can be said to be different from their classical counterparts.

This question can be addressed from many angles, and many traditional thermodynamic tasks have been studied in the quantum regime \cite{Levy2018}. The resources available to the machine will influence the answer, and it therefore seems particularly interesting to consider a minimal scenario in which no source of coherent control, work input or external driving is present and quantum thermal machines rely only on time-independent internal and system-environment interactions. Such machines are referred to as autonomous \cite{Mitchison2018, Mitchison2019}. Autonomous quantum thermal machines performing classical thermodynamic tasks have been explored e.g.~for heating \cite{Mahler2005, Roulet2017, Rosnagel2016, GelbwaserKlimovsky2018, Josefsson2018}, cooling \cite{Linden2010, Levy2012, Brunner2014,Clivaz2019, GelbwaserKlimovsky2014, Maslennikov2019}, heat management \cite{Senior2020, Iorio2021} and keeping time \cite{Erker2017, Woods2019}.

A complementary approach, which we pursue here, is to consider tasks that have no classical analogue; the paradigmatic example being that of entanglement generation. Interestingly, despite their simplicity, it turns out that autonomous machines are able to generate quantum correlations between initially independent systems. Two qubits and two out-of-equilibrium heat baths are sufficient to enable an entanglement engine that generates steady state entanglement \cite{Brask2015} by exploiting a heat current through the machine \cite{Khandelwal2020}. This can also be extended to multipartite systems \cite{Manyqubitengine}. In the simplest setting, the entanglement is weak and noisy, but it was found that stronger entanglement can be achieved in more sophisticated machines \cite{Man2019, Tavakoli2018, Tacchio2018, Wang2019}.  While these examples showcase the possibility of generating entanglement in quantum thermal machines and its nontrivial relationship to the resources made available, little is known about the fundamental limitations of such entanglement generation, the comparative power of different types of machines or their general ability to produce useful entanglement.

\begin{center}
\begin{figure*}[t!]
	\centering
	\includegraphics[width=2\columnwidth]{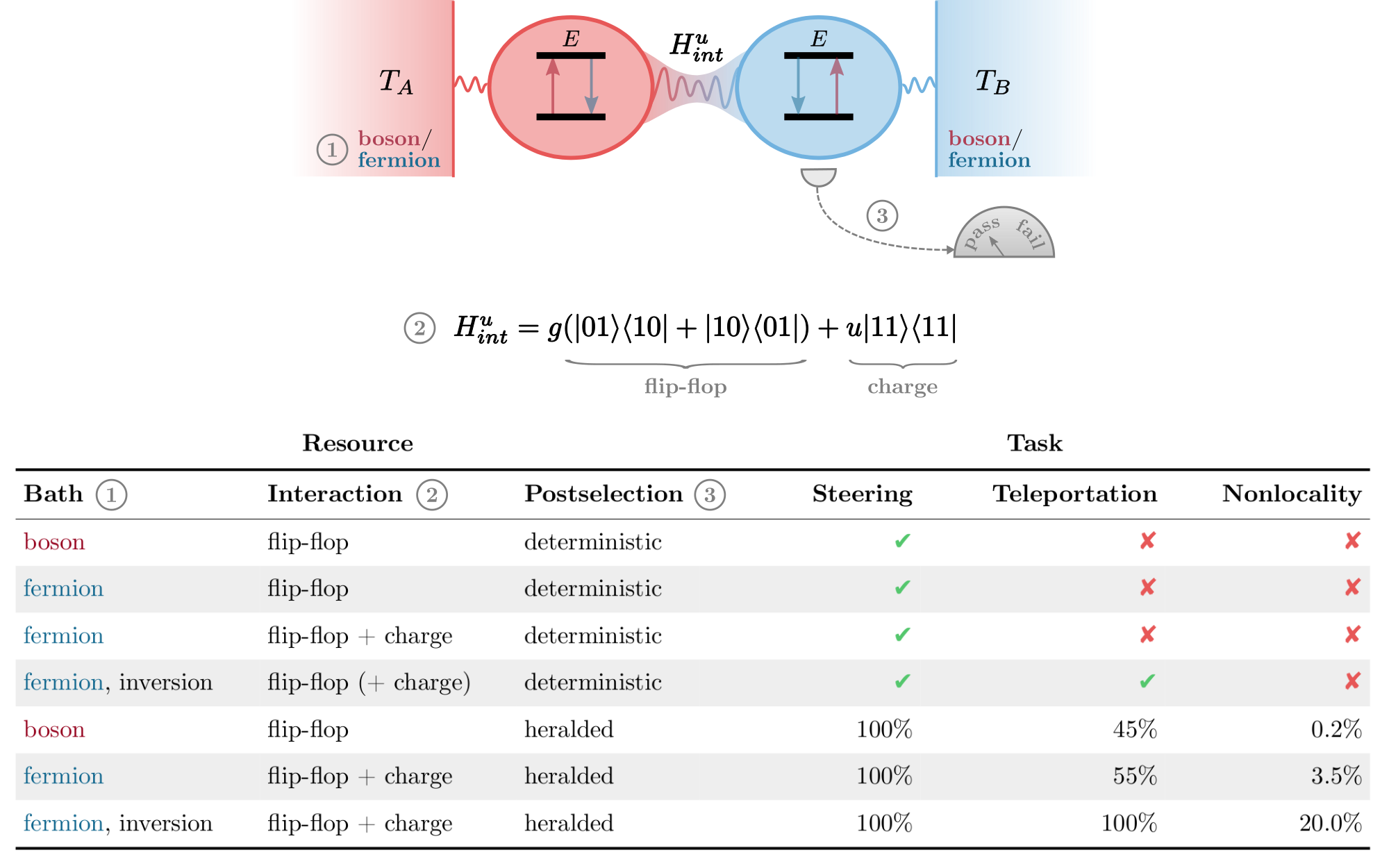}
	\caption{\textbf{(Top)} Two-qubit autonomous quantum thermal machine for generating entanglement. Two qubits with energy gap $E$ interact via a time-independent Hamiltonian and are individually coupled to thermal baths at temperatures $T_A$ and $T_B$. The heat current through the system drives the qubits into an entangled steady state. In this work, the baths may be bosonic or fermionic with or without population inversion, the interaction may be simple flip-flop or contain a `charge' term, and the entanglement is quantified either directly in the steady state or after postselection by local filtering on one or both qubits. \textbf{(Bottom)} Summary of results. We consider the generation of entanglement useful for steering, teleportation, or nonlocality for different combinations of bath type, interaction type, and postselection. In the deterministic setting, the right-hand side of the table shows whether such generation is possible. In the heralded setting, it shows a lower bound on the maximal heralding efficiency for which it is possible.}\label{TabMain}
\end{figure*}
\end{center}

Here, we systematically investigate entanglement generation in autonomous thermal machines. Our approach consists in analysing classes of machines that are characterised by exploiting qualitatively different resources. As resources, we consider combinations of different types of thermal baths, different qubit-qubit interactions, and the possibility to either utilise the steady state of the machine directly or to first apply local filtering.  These different resources are labeled 1-3 in Fig.~\ref{TabMain}. For the baths, we consider the two fundamental particle types, bosons and fermions. For the latter, we also consider the possibility of exploiting population inversion, i.e.~negative temperature baths (this is not relevant in the bosonic case, where the energy spectra are unbounded from above). For the interaction, we consider an energy-preserving flip-flop interaction, where the qubits can coherently exchange a single excitation, and we also allow for a direct term, increasing the energy of double-excited states. The latter can be thought of e.g.~as a Coulomb term for charged fermions. Finally, we also consider heralding using local filters diagonal in the local energy eigenbasis. Here, the amount entanglement conditioned on successful filtering is quantified. In contrast to previous works investigating heralded autonomous entanglement generation \cite{Tavakoli2018, Manyqubitengine}, our filtering process does not require higher-dimensional quantum systems and may be viewed as an add-on to the simplest machine.

Furthermore, in order to meaningfully quantify and compare autonomous entanglement generation, we adopt an operational approach. Previously, the quality of the entanglement produced in autonomous machines was primarily considered based on one of the well-known entanglement monotones, e.g.~entanglement concurrence  \cite{Hill1997, Wootters1998}  or entanglement negativity \cite{Vidal2002}. In the minimal, two-qubit, scenario these monotones can detect every entangled state and successfully identify the maximally entangled states. However, away from this extremal case, they typically do not reveal much knowledge about the nonclassicality of the state. For instance, there are states with an infinitesimal entanglement concurrence that can be used to violate a Bell inequality\footnote{For example, the pure family of states $\ket{\phi_{\alpha}}= \sqrt{\alpha} \ket{00} + \sqrt{1-\alpha} \ket{11}$ is entangled, and hence according to \cite{Gisin1991} Bell non-local, for all $\alpha \in (0,1)$. Its concurrence, however, is $C(\ket{\phi_{\alpha}})= \sqrt{4 \alpha (1-\alpha)}$, which can be arbitrarily small.} while other states with sizable entanglement concurrence cannot (e.g.\ noisy Werner states \cite{Werner1989}). More generally, entanglement measures only represent a partial order under local operations and classical communication \cite{Virmani2000}. In practice, this means e.g.\ that one state can be more entangled than another according to the concurrence, but vice versa according to the negativity \cite{Verstraete2001}. This motivates us to depart from these ideas and instead take an operational perspective in which entanglement is detected and quantified based on its ability to perform concretely useful tasks. To this end, we focus on three types of operational nonclassicality: Einstein-Podolsky-Rosen steering, quantum teleportation and Bell nonlocality. We choose these tasks as they are paradigmatic both in the foundations of quantum theory (see e.g.~the reviews \cite{Uola2020, Horodecki2009, Brunner2014b}) and in quantum information processing (see e.g.~\cite{Pirandola2015, Cavalcanti2016, Supic2020, Tavakoli2021}). Moreover, they also represent increasingly sophisticated notions of nonclassicality. Notably, steering and teleportation also enable other important forms of nonclassicality, namely quantum contextuality \cite{Tavakoli2020} and semi-device-independent quantum information entanglement detection \cite{FakeTriangle}. 

For the various classes of quantum thermal machines and the different notions of operationally useful entanglement, we obtain both constructive examples of nonclassicality and no-go results. The formers account for the ability of the machines to go beyond classical limitations and they allow for general comparisons between the performance of different quantum thermal machines. The latters represent fundamental limitations in the ability of quantum thermal machines to produce useful entanglement when granted particular resources. Together, these results provide insight both to the abilities and limitations of autonomous entanglement generation. Figure~\ref{TabMain} shows a sketch of the machines we consider and summarises our main results. For the most elementary machine, introduced in Ref.~\cite{Brask2015}, we surprisingly find that both bosonic and fermionic baths and interactions with and without the charge term all enable steering. This highlights the relevance of our operational approach: steering is possible in this machine but one can prove that the output state with the largest entanglement concurrence reported in Ref.~\cite{Brask2015} is, in fact, not steerable. However, we prove no-go theorems for both teleportation and a violation of the Clauser-Horne-Shimony-Holt (CHSH) Bell inequality: both are impossible with the most elementary machine regardless of bath and interaction type. Next, we show that, for fermionic baths, population inversion enables teleportation, but that it remains impossible to violate the CHSH inequality. Finally, we include local filtering and find that Bell nonlocality is then possible. In fact, such heralded entanglement generation enables the asymptotic generation of even the maximally entangled state. For bosons and charged fermions, teleportation is possible at high heralding efficiency. Moreover, when combined with  population inversion, we find that violations of the CHSH inequality can be sustained up to an efficiency of $20\%$. This enables efficient generation of the strongest form of nonclassicality from thermal resources. 

The remainder of the paper is structured as follows. In Sec.~\ref{secScenario}, we describe the two-qubit thermal machines, which we consider. In Sec.~\ref{secEnt}, we define the operational measures of nonclassicality, which we use to quantify the usefulness of the states generated by the machines. In Sec.~\ref{secSimpleM}, we study the simplest machines, which do not use population inversion nor postselection, in Sec.~\ref{secnegtemp}, we allow for inversion for machines with fermionic baths, and in Sec.~\ref{secHerald} we further allow for local filtering, resulting in non-deterministic, heralded machines. Finally, in Sec.~\ref{secConc} we conclude.

\section{Scenario}\label{secScenario}
Throughout, we adopt natural units in which Planck's constant and Boltzmann's constant are both unity, i.e.~$\hbar=k_\text{B}=1$. Our machine consists of two resonant qubits, labeled A and B, with excited state energy $E$ and free Hamiltonians $H_\A=E\ketbra{1}{1}_\A \otimes \openone_\B$ and $H_\text{B}= \openone_\A\otimes E\ketbra{1}{1}_\B$, where we have set the ground-state energy to zero. For simplicity, we normalise the excited state energy to $E=1$.  The qubits interact via a time-independent Hamiltonian of the form 
\begin{equation}
H_\text{int}^{u}=g\left(\ketbra{01}{10}+\ketbra{10}{01}\right)+u\ketbra{11}{11} .
\end{equation}
Here, the first term describes a flip-flop interaction with strength $g\geq 0$, while the second term, with strength $u\geq0$, increases the energy of the double-excited state $\ket{11}$. We refer to this term as the `charge' term, as it would e.g. arise from the Coulomb repulsion between electrons in charged-quantum-dot qubits, as considered in \cite{Brask2015}. Thus, for `uncharged' qubits, we set $u=0$ whereas for `charged' qubits we have $u>0$. Each qubit, $\A$ and $\B$, is individually coupled to a bath of temperature $\TA$ and $\TB$ respectively (see the top part of Fig.~\ref{TabMain}) with coupling strength $\gamma_\A$, $\gamma_\B$ respectively. 
Under a weak system-bath coupling with respect to the bare energies of the qubits, $\gamma_{\A},\gamma_\B \ll E$, the dynamics of the two-qubit state $\rho=\rho_\text{AB}$ can be described with a Lindblad master equation. Throughout this work, we assume in addition that the inter-qubit coupling strength $g$ remains smaller or of the order of $\gamma_{\A}$ and $\gamma_\B$, we work in the parameters' range $g \leq (\gamma_{\A},\gamma_\B) \ll E, u$. This parameters range ensure that dissipation to the environments is correctly described by local jump operators, i.e.~jump operators acting locally onto each qubit separately \cite{Purkayastha2016, Hofer2017, Gonzalez2017, Mitchison2018, DeChiara2018, Cattaneo2019}. In the case of a strong inter-qubit coupling, the possibility of thermal-state entanglement is not precluded (ground state can be entangled). In Ref.~\cite{Khandelwal2020}, it was shown that the strong-coupling regime does not provide any advantage with respect to the weak inter-qubit coupling regime for the creation of entanglement in an out-of-equilibrium situation. We now discuss the exact form of the local master equation for the different models we consider.

\textbf{Bosons without charge.} For bosons, the bath statistics is described by a Bose-Einstein distribution $n_\text{BE}(\varepsilon,T)=\left(e^{\frac{\varepsilon}{T}}-1\right)^{-1}$. Assuming a weak system-bath coupling strength $\gamma$, the rate of receiving excitations at energy $\varepsilon$ from a bath at temperature $T$ is then given by $\Gamma^+_{\text{BE}}(\gamma,\varepsilon,T)=\gamma n_\text{BE}\left(\varepsilon,T\right)$, and the rate of loosing excitations into the bath is $\Gamma^-_{\text{BE}}(\gamma,\varepsilon,T)=\gamma\left(1+n_\text{BE}\left(\varepsilon,T\right)\right)$ \cite{Breuertextbook}. For bosons, we will consider only the uncharged interaction ($u=0$). In this case, excitations can only be lost or gained at energy $E$ and the dynamics of the two-qubit state is then described by a Lindblad master equation
\begin{multline}\label{lindbladBoson}
\frac{\partial \rho}{\partial t}=i\left[\rho,H_\text{tot}\right]+\sum_{\substack{k\in\{\A,\B\}}}\Gamma_{k}^+\left(J_{k}\rho J_{k}^\dagger-\frac{1}{2}\{J_{k}^\dagger J_{k},\rho\}\right)\\
+\sum_{\substack{k\in\{\A,\B\}}}\Gamma_{k}^-\left(J_{k}^\dagger\rho J_{k}-\frac{1}{2}\{J_{k} J_{k}^\dagger,\rho\}\right),
\end{multline}
where $H_\text{tot}=H_\A+H_\B+H_\text{int}^0$ is the total system Hamiltonian, the rates are $\Gamma_k^\pm = \Gamma^\pm_{\text{BE}}(\gamma_k,E,T_k)$ with $\gamma_k$ the strength of the coupling to bath $k$, and the local jump operators are $J_{\A}= \ketbra{1}{0}\otimes \openone$ and $J_{\B}= \openone \otimes \ketbra{1}{0}$.

\textbf{Fermions without charge.} For fermionic baths, an uncharged interaction and the same parameters' range $g \leq (\gamma_{\A},\gamma_\B) \ll E$, the dynamics of the system is again described by a local master equation \eqref{lindbladBoson} but with the modification that the bath statistics is now given by the Fermi-Dirac distribution $n_\text{FD}(\varepsilon,T)=\left(1+e^{\frac{\varepsilon}{T}}\right)^{-1}$. The rates of receiving or loosing excitations are then given by $\Gamma_{FD}^+(\gamma,\varepsilon,T)=\gamma n_\text{FD}\left(\varepsilon,T\right)$ and $\Gamma_{FD}^-=\gamma\left(1-n_\text{FD}\left(\varepsilon,T\right)\right)$ respectively, and in \eqref{lindbladBoson} one has $\Gamma^\pm_k = \Gamma^\pm_{FD}(\gamma_k,E,T_k)$. 

\textbf{Fermions with charge.} For fermionic baths we also consider the possibility of a charged interaction (i.e.~$u>0$). In this case, when both qubits are excited, the total energy is increased to $2E+u$. Dissipative transitions can therefore occur both at energy $E$ and at energy $E+u$, and the master equation must be modified accordingly. Defining the rates  $\Gamma_{kl}^+=\gamma_k n_\text{FD}\left(E+lu,T_k\right)$ and $\Gamma_{kl}^-=\gamma_k\left(1-n_\text{FD}\left(E+lu,T_k\right)\right)$, the Lindblad equation takes the form  
\begin{multline}\label{lindblad}
\frac{\partial \rho}{\partial t}=i\left[\rho,H_\text{tot}\right]+\sum_{\substack{k\in\{\A,\B\}\\ l\in\{0,1\}}}\Gamma_{kl}^+\left(J_{kl}\rho J_{kl}^\dagger-\frac{1}{2}\{J_{kl}^\dagger J_{kl},\rho\}\right)\\
+\sum_{\substack{k\in\{\A,\B\}\\ l\in\{0,1\}}}\Gamma_{kl}^-\left(J_{kl}^\dagger\rho J_{kl}-\frac{1}{2}\{J_{kl} J_{kl}^\dagger,\rho\}\right),
\end{multline}
where the total system Hamiltonian is now $H_\text{tot}=H_\A+H_\B+H_\text{int}^u$ and the jump operators are $J_{\A 0}=\ketbra{1,0}{0,0}$, $J_{\A 1}=\ketbra{1,1}{0,1}$, $J_{\B 0}=\ketbra{0,1}{0,0}$ and $J_{\B 1}=\ketbra{1,1}{1,0}$. Here, the operators $J_{\A0}$ and $J_{\B0}$ correspond to transitions at energy $E$ and the  operators $J_{\A1}$ and $J_{\B1}$ correspond to transitions at energy $E+u$.


At long times, the dynamics of both \eqref{lindbladBoson} and \eqref{lindblad} converges to a steady state $\rho_\text{steady}$, which is the solution to $\frac{\partial \rho}{\partial t}=0$. The steady state is determined completely by the machine parameters $(g,\gA,\gB,\TA,\TB,u)$ and its exact form is in general complicated. However, in the minimal model we consider, made of two qubits interacting via a flip-flop type Hamiltonian, the only non-vanishing off-diagonal element in the steady state will be the one sustained by the unitary evolution, hence by the interaction Hamiltonian that induces exchanges between the states $\ket{01}$ and $\ket{10}$ in the basis of the free Hamiltonians. Therefore, in all the cases we consider, the solution is of the form
\begin{equation}\label{form}
\rho_\text{steady}=\begin{pmatrix}
a_1 &0&0&0\\
0& a_2 & -\alpha & 0\\
0&-\alpha^* & a_3&0\\
0&0&0&1-a_1-a_2-a_3 
\end{pmatrix},
\end{equation}
for some $(a_1,a_2,a_3)\geq 0$ such that $|\alpha|\leq \sqrt{a_2a_3}$ (positivity) and $a_1+a_2+a_3\leq 1$ (normalisation).  Notably, when investigating the entanglement properties of \eqref{form}, it is convenient to restrict  $\alpha$ to be a real, non-negative number. This comes at no loss of generality since entanglement (as well as steering, teleportation and nonlocality) is invariant under local unitary operations. Specifically, if $\alpha=|\alpha|e^{i\bar{\alpha}}$, where $\bar{\alpha}$ is a phase, then we can apply the unitary $U=\ketbra{0}{0}+e^{i\bar{\alpha}}\ketbra{1}{1}$ to qubit A in order to transform the state \eqref{form} into
\begin{equation}\label{form2}
\begin{pmatrix}
a_1 &0&0&0\\
0& a_2 & -|\alpha| & 0\\
0&-|\alpha|& a_3&0\\
0&0&0&1-a_1-a_2-a_3 
\end{pmatrix},
\end{equation}
In the following section, where we formulate criteria for nonclassicality, we focus on states of the form \eqref{form2}.


\section{Operationally useful entanglement}\label{secEnt}

We quantify the nonclassicality of the quantum correlations in $\rho_\text{steady}$ by their usefulness for three operational tasks, namely steering, teleportation and nonlocality. In this section, we discuss each of them separately.

\subsection{Steering}\label{secsteering}

Consider that one qubit is given to Alice and the other qubit to Bob. By performing different projective measurements $\{A_{a|x}\}$, where $x$ is the setting and $a$ the outcome, Alice remotely prepares (unnormalised) states  $\sigma_{a|x}=\Tr_\text{A}\left(A_{a|x}\otimes \openone\rho\right)$ for Bob. The set $\{\sigma_{a|x}\}$ is called an assemblage and it is said to be steerable if it does not admit a local-hidden-state model. This means that it cannot be explained by a model in which Alice uses a local random variable, $\lambda$, to stochastically prepare quantum states, $\sigma_\lambda$, for Bob and chooses her output via local post-processing \cite{Wiseman2007}: $\sigma_{a|x}=\sum_\lambda p(\lambda) p(a|x,\lambda)\sigma_\lambda$. The state $\rho$ is steerable if there exists local measurements for Alice that generate a steerable assemblage. In contrast, if a local hidden state model for the assemblage exists for every possible set of projective measurements, then   $\rho$ is said to be unsteerable. 

The (un)steerability of a given assemblage, obtained from $N$ projective measurements with $O$ possible outcomes each, can be determined through the following semidefinite program
\begin{align}\label{steeringsdp}\notag
&\text{find } \{\bar{\sigma}_\lambda\} \quad \text{such that} \quad \bar{\sigma}_\lambda\geq 0, \\
&\forall (a,x): \hspace{3mm}\sigma_{a|x}=\sum_{\lambda=1}^{O^{N}}D(a|x,\lambda)\bar{\sigma}_\lambda
\end{align}
where $D(a|x,\lambda)$ are all possible deterministic probability distributions of Alice. The assemblage is steerable if and only if a solution to \eqref{steeringsdp} cannot be found. This semidefinite program becomes more expensive to evaluate as the number of settings and outcomes increases. In the interest of a reasonable compromise, we will often consider detecting steerability based on Alice performing a set of ten qubit projective measurements whose Bloch vectors form a regular dodecahedron on the Bloch sphere. Such a dodecahedral measurement configuration has been found useful in previous works for detecting quantum correlations \cite{Saunders2010, Plato}.

However, it is less straightforward to assert the unsteerability of a given state. For states of two qubits, computable sufficient conditions (in the form of linear programs) for steerability and unsteerability respectively, under any number of projective measurements, were provided in Ref.~\cite{Nguyen2019}. These are based on approximating the Bloch sphere from both the inside and outside with polytopes of increasing number of vertices in order to obtain increasingly precise upper and lower bounds on the so-called critical steering radius, which is necessary and sufficient to determine the steerability of general two-qubit states (when no limitation is placed on the number of measurements) \cite{Nguyen2019}. 

Moreover, since some states display stronger steering properties than others, one may consider one of the many possible quantifiers of steering  \cite{Cavalcanti2016}. Here, we will focus on the amount of isotropic (white) noise that a state can undergo without becoming unsteerable. This amounts to substituting a two-qubit state according to $\rho \rightarrow (1-q)\rho_\text{steady}+q\frac{\openone}{4}$ where $q\in[0,1]$ is the noise rate, and finding an upper bound on $q$ below which the state remains steerable.

Lastly, we note that every steerable state enables a proof of another type of quantum correlations, namely quantum contextuality \cite{Tavakoli2020}. Therefore, if a state is steerable, it is also able to produce outcome statistics that elude classical models in contextuality experiments.

\subsection{Teleportation} Quantum teleportation \cite{Teleportation} allows Alice to send Bob a qubit state $\ket{\psi}$ by means of shared entanglement $\rho$ and classical communication. In a standard teleportation protocol, she jointly projects $\ket{\psi}$ and her part of $\rho$ in a basis of four maximally entangled states and sends the outcome to Bob. Bob applies a local unitary depending on the outcome. In an ideal setting, in which $\rho $ is maximally entangled, Bob then recovers $\ket{\psi}$ exactly. However, also non-maximally entangled states enable a degree of teleportation. The standard quantifier for the task is the fidelity of teleportation, $f$, which measures the closeness between Bob's final state and $\ket{\psi}$. It is given by $f=\frac{1+2F}{3}$ where 
\begin{equation}\label{SF}
F(\rho)=\max_U \bracket{\psi^-}{\left(\openone\otimes U\right)\rho \left(\openone\otimes U^\dagger\right)}{\psi^-},
\end{equation} 
is the so-called singlet fraction, where $\ket{\psi^-}=\frac{\ket{01}-\ket{10}}{\sqrt{2}}$ and $U$ is a qubit unitary.  Since the best classical protocol achieves $f=\frac{2}{3}$, the state $\rho$ is useful for teleportation if and only if $F>\frac{1}{2}$ \cite{Horodecki1999}. Therefore, when $F> \frac{1}{2}$, we say that the singlet fraction is nontrivial.

In general, there is no closed  expression for $F$ but when restricting to states of the form \eqref{form2}, we derive a simple expression. A general qubit unitary is written $U=e^{i\mu \vec{n}\cdot \vec{\sigma}}$ where $\vec{n}=\left(\sin\theta\cos\phi,\sin\theta\sin\phi,\cos\theta\right)$ is some Bloch vector, $\mu$ is a rotation angle and $\vec{\sigma}=(\sigma_x,\sigma_y,\sigma_z)$ is a vector of Pauli matrices. By direct calculation, one finds that the expression $\bracket{\psi^-}{(\openone\otimes U)\rho \left(\openone\otimes U^\dagger\right)}{\psi^-}$ has no dependence on $\phi$ and that its optimal value is achieved for $\theta=\frac{\pi}{2}$ if $\left(1+2\alpha-2\Delta\right)\geq0$ and for $\theta=0$ otherwise, where we defined $\Delta=a_2+a_3$, see \eqref{form2}. Similarly, one finds that the optimal choice of $\mu$ is either $\mu=0$ or $\mu=\frac{\pi}{2}$. This leads to
\begin{equation}\label{singletfrac}
F=\begin{cases}
\alpha+\frac{\Delta}{2} & \text{if } \left(1+2\alpha-2\Delta\right)\leq 0\\
\max\{\alpha+\frac{\Delta}{2},\frac{1-\Delta}{2}\} & \text{otherwise}
\end{cases}.
\end{equation}
Consequently, if the singlet fraction is nontrivial, it is equal to $\alpha+\frac{\Delta}{2}$. A necessary and sufficient condition for a trivial singlet fraction becomes
\begin{equation}\label{telecond2}
\alpha+\frac{\Delta}{2}\leq \frac{1}{2} \quad\Leftrightarrow \quad F\leq \frac{1}{2}  
\end{equation}
A simpler, sufficient but not necessary condition for a trivial singlet fraction is obtained by invoking positivity of the density matrix, namely $\alpha\leq \sqrt{a_2a_3}$ in \eqref{singletfrac}. Then, we have that 
\begin{equation}\label{telecond}
\Delta\leq \frac{1}{2} \quad\Rightarrow \quad F\leq\frac{1}{2}.
\end{equation}

Lastly, it is relevant to note that every two-qubit state that is useful for teleportation also enables semi-device-independent entanglement certification through an experiment in the spirit of quantum dense coding \cite{FakeTriangle}. Therefore, an advantage in teleportation implies this alternative notion of quantum correlations.

\subsection{Nonlocality}

Consider that Alice and Bob each perform measurements $x$ and $y$ on the shared state $\rho$ and record their outcomes $a$ and $b$. The resulting probability distribution $p(a,b|x,y)$ is said to be nonlocal if it cannot be explained by a local-hidden-variable model, in which Alice and Bob decide each their respective outcomes based on a shared random variable $\lambda$ \cite{Bell}: $p(a,b|x,y)=\sum_\lambda p(\lambda) p(a|x,\lambda)p(b|y,\lambda)$. Nonlocality is therefore a stronger form of nonclassicality than steering. Here, we focus exclusively on the the simplest and most popular Bell scenario. It has binary inputs and outputs, $a,b,x,y\in\{0,1\}$, and is fully characterised by the CHSH inequality \cite{CHSH},
\begin{equation}\label{chsh}
\text{CHSH}=\sum_{x,y,a,b}(-1)^{a+b+xy}p(a,b|x,y)\leq 2.
\end{equation}
Quantum theory can produce a violation of at most $\text{CHSH}=2\sqrt{2}$, which is achieved with a maximally entangled state \cite{Cirelson1980}.

Given an arbitrary two-qubit state $\rho$, the largest CHSH-value (for the best pairs of local measurements) can be determined via the Horodecki criterion \cite{horodeckicriterion}. This criterion stipulates that the largest value is $\text{CHSH}=2\sqrt{\lambda_1+\lambda_2}$ where $\lambda_1$ and $\lambda_2$ are the two largest eigenvalues of the operator $T^\text{T}T$, where $T_{ij}=\Tr\left(\rho \sigma_i\otimes \sigma_j\right)$ for $i,j\in\{x,y,z\}$. For states of the form \eqref{form2}, this operator simplifies into $T=\text{diag}(-2\alpha,-2\alpha,1-2\Delta)$. Hence, the eigenvalues of $T^\text{T}T$ are $\left(4\alpha^2,4\alpha^2,(2\Delta-1)^2\right)$, leading to  
\begin{equation}
\text{CHSH}=2\sqrt{8\alpha^2+(2\Delta-1)^2-\min\{4\alpha^2,(2\Delta-1)^2\}}.
\end{equation}
A handy upper bound on the CHSH parameter is obtained by discarding the min-term, which leads to the following sufficient condition for satisfying the CHSH inequality
\begin{equation}\label{chshcond}
8\alpha^2+(2\Delta-1)^2\leq 1 \quad \Rightarrow \quad \text{CHSH}\leq 2.
\end{equation}
A useful, but weaker, sufficient condition is obtained from \eqref{chshcond} by invoking positivity ($\alpha\leq \sqrt{a_2 a_3}$) and the fact that the polynomial $a_2^2+a_3^2+4a_2a_3-a_2-a_3$ is non-positive in the domain $(a_2,a_3)\geq 0$ and $a_2+a_3\leq \frac{1}{2}$. One obtains
\begin{equation}\label{chshcond2}
\Delta\leq \frac{1}{2} \quad\Rightarrow \quad \text{CHSH}\leq 2.
\end{equation}
Note that \eqref{chshcond2} is identical to the condition \eqref{telecond} for inability of teleportation.

\section{Nonclassicality in the simplest machine}\label{secSimpleM}
In this section we investigate the simplest autonomous entanglement engine, i.e.~the machine introduced in section \ref{secScenario}, for the case of bosonic and fermionic baths and uncharged flip-flop interaction, and for fermionic baths and charged flip-flop interaction. For each of these cases, we consider the possibility of generating steady state entanglement that is strong enough to enable steering, teleportation and nonlocality. In all cases, we find that steering is possible, while teleportation and a violation of the CHSH inequality are not.

\begin{figure}
	\centering
	\includegraphics[width=\columnwidth]{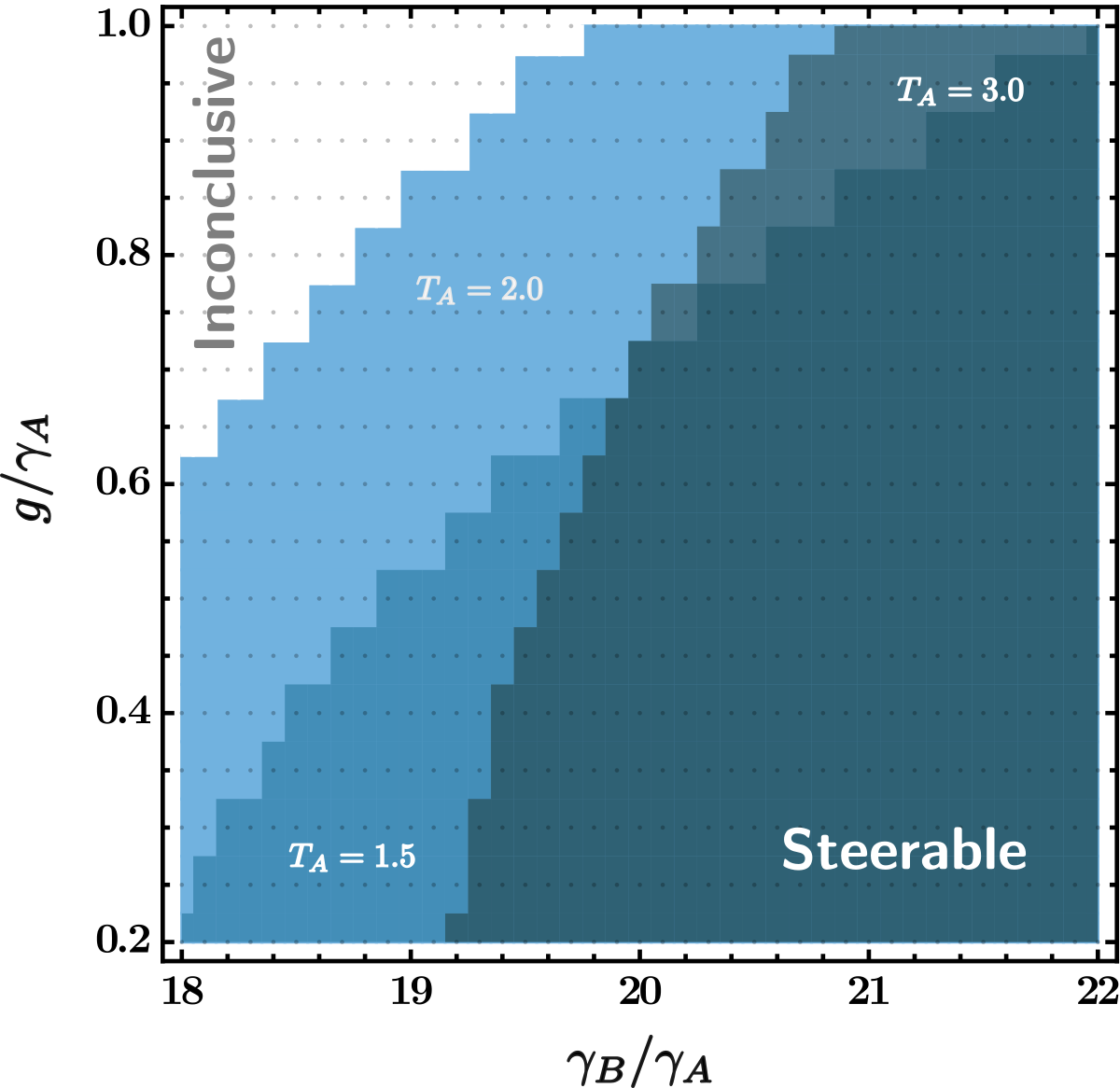}
	\caption{Steerability of steady state, in the space $\left(\frac{g}{\gA},\frac{\gB}{\gA}\right)$ for different choices of $\TA$, using any number of projective measurements. The range for the different parameters was chosen to ensure the validity of our master equations and in accordance with state-of-the-art experiments, see \cite{Brask2015, Khandelwal2020}. The plot is generated using the linear programming methods of Ref.~\cite{Nguyen2019} to bound the critical steering radius. The dots represent sampled points and the shaded regions represent a steerable steady state. The white region represents points for which we are unable to determine whether the steady state is steerable or not.}\label{FigBosonSteering}
\end{figure}

\subsection{Bosons}
When the baths are bosonic, we solve for the steady state version of \eqref{lindbladBoson}. This can be achieved analytically by listing the density-matrix variables in \eqref{form} in a vector $\vec{x}$ and solving an inhomogeneous system of linear equations, $0=A\vec{x}+b$, where the matrix $A$ and the vector $b$ depend on the machine parameters $(g,\gA,\gB, \TA,\TB)$. See \cite{Khandelwal2020} for an explicit derivation. In Appendix~\ref{AppBosons}, we present the analytical expression for $\rho_\text{steady}$ in the most relevant limit, namely $T_\B\rightarrow 0$. It is notable that the steady state does not explicitely depend on the parameters $ (g,\gA,\gB)$ but instead only on the ratios $g/\gA$ and $\gB/\gA$. Thus, the steady state remains the same if all parameters $ (g,\gA,\gB)$ are multiplied by a positive constant (see Appendix.~\ref{AppBosons}). In fact, since this property occurs frequently, we will exploit it several times in the manuscript to fix, without any loss of generality, $\gA=1$. Throughout the work, we ensure that all parameters remain in a range satisfying the conditions for assessing the dynamics with a local Lindblad master equation as discussed above.

By inspecting the steady state, we show in Appendix~\ref{AppBosons} that the condition \eqref{telecond} is satisfied for every choice of machine parameters. Therefore, this machine cannot produce entanglement strong enough for teleportation. Moreover, by virtue of condition \eqref{chshcond2}, it immediately follows that there exists no steady state that can violate the CHSH inequality.

However, there exist choices of machine parameters for which the corresponding steady state becomes steerable from Alice to Bob for sufficiently many projective measurements. To investigate this, we have employed the linear programming relaxations of Ref.~\cite{Nguyen2019} for bounding the critical steering radius (see section~\ref{secsteering}). In Figure~\ref{FigBosonSteering}, we illustrate the results of a grid-search for steerability in the space of the steady state, characterised by the ratios $\left(\frac{g}{\gA},\frac{\gB}{\gA}\right)$, at different choices of temperature $\TA$. We consider a finite inter-qubit coupling strength $g$ to ensure the presence of quantum coherence in the steady state. We see that different choices of $\TA$ correspond to different parameter regions in which the steady state is steerable. Notably, however, the steerability is very fragile, in the sense that a small amount of isotropic noise is sufficient to enable a local hidden state model. For such reasons, we have also been unable to detect steerability using the dodecahedral measurement configuration for Alice (see section~\ref{secsteering}). This suggests that one may need a sizable number of projective measurements in order to generate a steerable assemblage from the steady state.

\subsection{Fermions without charge}
For fermionic baths and uncharged interaction, we have again solved the master equation \eqref{lindbladBoson} as a function of the machine parameters by means similar to the above discussed case of bosons. For fermions, due to the Fermi-Dirac statistics, the most interesting scenario emerges in the limit of a large temperature gradient, i.e.~when $\TA\rightarrow \infty$ and $\TB\rightarrow 0$. In this limit, the steady state takes the form

\begin{widetext}
\begin{equation}\label{minimalsteady}
\rho_\text{steady}=\frac{1}{2N}\begin{pmatrix}
& \gA\gB t^2+2g^2 s^2 & 0 &0&0\\
& 0 & 2g^2\gA s & -2igt\gA\gB & 0\\
&0 & 2itg\gA\gB & \gA\left(\gB t^2+2g^2 s\right) & 0\\
& 0 & 0&0& 2g^2\gA^2
\end{pmatrix},
\end{equation}
\end{widetext}
where $t=\gA+\gB$, $s=\gA+2\gB$, and $N=t^2\left(4g^2+\gA\gB\right)$. This state always satisfies the condition \eqref{telecond}, which asserts that it can neither enable teleportation nor a violation of the CHSH inequality (see condition \eqref{chshcond2}). To prove this, we consider the sum $\Delta$ of the second and third diagonal element in \eqref{minimalsteady}. One easily finds that the derivative of $\Delta$ w.r.t.~$g$ is non-positive and equals zero only at $g=0$. In this limit, we have $\Delta\leq \frac{1}{2}$.

\begin{center}
\begin{figure*}
	\centering
	\includegraphics[width=2\columnwidth]{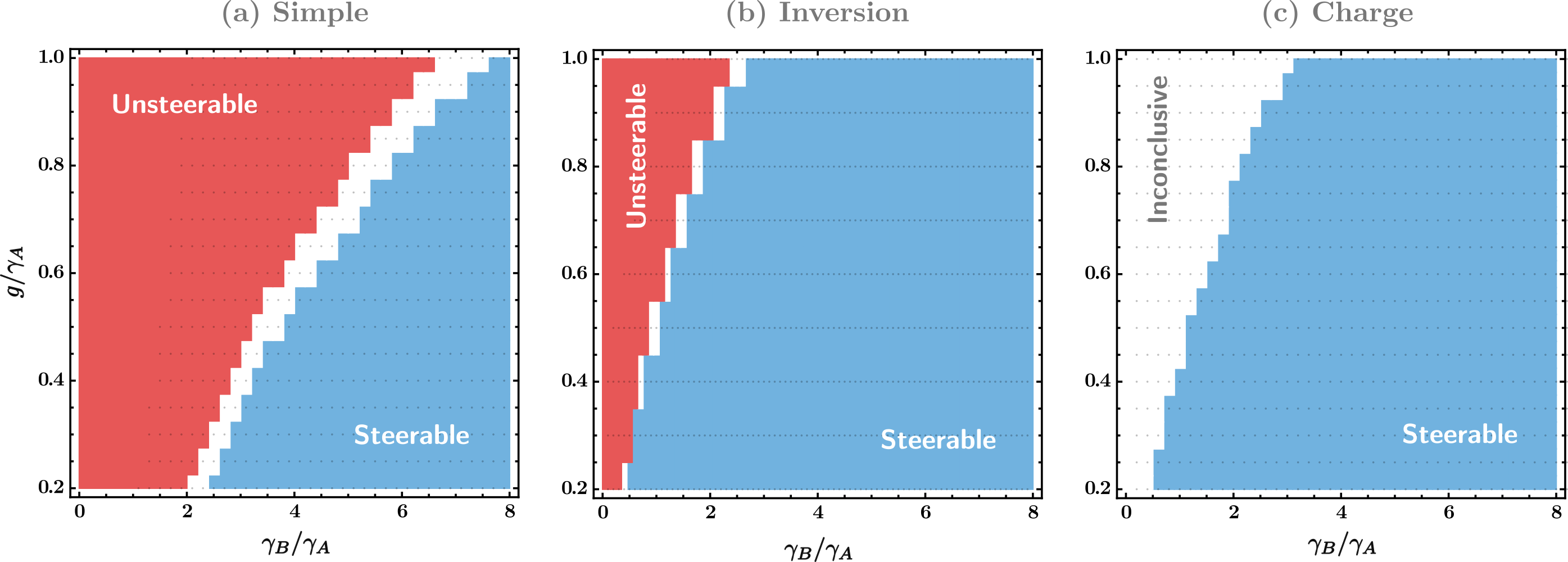}
	\caption{Steerability of steady state in the space of machine parameters using any number of projective measurements. The range for the different parameters was chosen to ensure the validity of our master equations and according to state-of-the-art experiments, see \cite{Brask2015, Khandelwal2020}. The plots are generated using the linear programming methods of Ref.~\cite{Nguyen2019} to bound the critical steering radius. The dots represent sampled points, and the blue (red) region represents (un)steerable steady states. The white region represents points for which we are unable to determine whether the steady state is steerable or not. (a) Simplest machine, fermions without charge, based on the steady state \eqref{minimalsteady}. (b) Fermions with population inversion; based on the steady state \eqref{steady}. (c) Fermions with charge; based on the steady state \eqref{steady2} in the most relevant limit of $\TA\rightarrow \infty$.}\label{Figsteeringfermion}
\end{figure*}
\end{center}

However, in analogy with the bosonic case, the steady state \eqref{minimalsteady} enables steering. As discussed previously, it suffices to scan the parameter range through the two ratios $\frac{g}{\gA}$ and $\frac{\gB}{\gA}$. We have conducted a grid search in the corresponding two-dimensional parameter space in order to determine the parameter region in which steering is (im)possible. For this qualitative assessment of steerability, we have used the linear programming method of Ref.~\cite{Nguyen2019} and the results are illustrated in Figure~\ref{Figsteeringfermion}a. We find that fermions enable steering at a much smaller value of $\frac{\gB}{\gA}$ than bosons. In addition, we have also quantitatively investigated the steering properties of the steady state, specifically through considering its robustness to isotropic noise, using a small number of measurements. To this end, we have employed the dodecahedral measurement configuration and at best found that a noise rate of $q\approx 0.23\%$ can be tolerated. This corresponds to choosing $\frac{g}{\gA}=0.6$ and  $\frac{\gB}{\gA}=8.5$. Although this noise tolerance is very small, it still improves on the case of bosons where such a dodecahedral measurement configuration did not detect steering.

\subsection{Fermions with charge}
The entanglement generated using the charged interaction is expected to be at least as powerful as that of the uncharged case. The reason is that the former is an immediate generalisation of the latter corresponding to allowing any non-zero value $u> 0$. Again, the most interesting case is that of a cold bath  $\TB\rightarrow 0$. The strongest entanglement is obtained in the limit of a large interaction, i.e.~when $u\rightarrow \infty$. The reason is that this completely suppresses population in the $\ket{1,1}$ subspace in the steady state, which is desirable since such a population effectively constitutes a form of noise due to the fact that the coherence of the state is concentrated in the subspace spanned by $\{\ket{01},\ket{10}\}$, see \eqref{form}. The steady state solution of the Lindblad equation \eqref{lindblad} can be expressed on the form \eqref{form} with coefficients 

\begin{widetext}
\begin{align}\label{steady2}
&a_1=\frac{\left(4g^2\left(1+e^\frac{1}{\TA}\right)+\gA\gB e^\frac{1}{\TA}\right)\left(\gB+t e^\frac{1}{\TA}\right)}{\left(1+e^\frac{1}{\TA}\right)\left(4g^2\left(\gA\left(2+e^\frac{1}{\TA}\right)+\gB\left(1+e^\frac{1}{\TA}\right)\right)+\gA\gB\left(\gB+te^\frac{1}{\TA}\right)\right)}\\
&a_2=\frac{4g^2\gA}{4g^2\left(\gA\left(2+e^\frac{1}{\TA}\right)+\gB\left(1+e^\frac{1}{\TA}\right)\right)+\gA\gB\left(\gB+te^\frac{1}{\TA}\right)}\\
& a_3=1-a_1-a_2\\
& \alpha=\frac{2ig\gA\gB}{4g^2\left(\gA\left(2+e^\frac{1}{\TA}\right)+\gB\left(1+e^\frac{1}{\TA}\right)\right)+\gA\gB\left(\gB+te^\frac{1}{\TA}\right)}.
\end{align}
\end{widetext}
Note that the strongest entanglement is expected in the limit $\TA\rightarrow \infty$.

In Appendix~\ref{AppColoumb} we prove that for any choice of machine parameters, when transformed into the form \eqref{form2}, this steady state satisfies the condition \eqref{telecond2}. Consequently, it cannot be used for teleportation. In the same Appendix, we prove that also condition \eqref{chshcond} is satisfied. Thus, it also cannot be used to violate the CHSH inequality. This constitutes a first no-go theorem, stating that neither teleportation nor nonlocality can be achieved with the simplest machine.

Since uncharged fermions already were found to enable steering, it trivially follows that charged fermions also enable steering. However, we find that the machine parameter region in which steering is possible is enlarged as compared to the case of uncharged fermions (see Figure~\ref{Figsteeringfermion}c). Moreover, the steerability is somewhat more robust to isotropic noise. To showcase this, we have considered the dodecahedral measurement configuration for Alice and found at best $q\approx 0.75\%$ at $\frac{g}{\gA}=0.3$ and $\frac{\gB}{\gA}=2.1$.

\section{Nonclassicality in machine with population inversion}\label{secnegtemp}
The inability to enable teleportation or nonlocality in the simplest autonomous machines motivates us to explore whether such nonclassicality can be achieved by supplementing the machine with the additional resource of population inversion. Specifically, we focus on fermionic systems and consider that the hot bath (bath A) is subjected to a population inversion process, analogous to that in a laser,  which allows the bath states to have a larger population in the excited state than in the ground state. This effectively corresponds to introducing a negative temperature in the Fermi-Dirac statistics of the bath, i.e.~we allow for $\TA<0$ which implies $\frac{1}{2}< n_\text{FD}\leq 1$. We will show that this resource enables teleportation, but not a violation of the CHSH inequality.

The idealised scenario, in which the inversion resource becomes most pronounced, is when the bath B is cold $\TB\rightarrow 0^+$ and bath A has a vanishing, negative, temperature, $\TA\rightarrow 0^-$. The latter corresponds to complete inversion, i.e.~$n_\text{FD}=1$. Notice that these limits eliminate any dependence on charge term ($u$) in the interaction Hamiltonian, which can be seen from examining the flow rates $\Gamma^\pm$ (see section~\ref{secScenario}). Therefore, we need not distinguish between charged and uncharged fermions. We have solved the steady state case of the  Lindblad equation \eqref{lindblad} and obtained the following solution:
\begin{equation}\label{steady}
\rho_\text{steady}=\frac{1}{N}\begin{pmatrix}
	4g^2\gB^2 & 0&0&0\\
	0& 4g^2\gA\gB & -2itg\gA\gB &0\\
	0 & 2itg\gA\gB & \gA \gB\left(4g^2+t^2\right)&0\\
	0&0&0& 4g^2\gA^2
\end{pmatrix}.
\end{equation}
Next, we investigate the usefulness of this steady state for teleportation, nonlocality and steering.

\subsection{Teleportation is possible}
We show through an explicit example that there exists a choice of machine parameters that corresponds to a steady state \eqref{steady} which is useful for teleportation. To this end, we choose $\gA=\gB$ and $g=\frac{\sqrt{5}-1}{4}\gB$. Using \eqref{singletfrac}, we evaluate the singlet fraction of the corresponding steady state to be $F(\rho_\text{steady})=\frac{3+\sqrt{5}}{8}\approx 0.65$, which exceeds the classical limit $F=\frac{1}{2}$. Equivalently, we may say that the teleportation fidelity is $f=\frac{7+\sqrt{5}}{12}\approx 0.77$, which exceeds the classical limit $f=\frac{2}{3}$.

More generally, whenever the singlet fraction is nontrivial, we can compute it from \eqref{singletfrac} and the steady state \eqref{steady}:
\begin{equation}\label{fidneg}
F=\frac{\gA\gB\left(8g^2+4gt+t^2\right)}{2t^2\left(4g^2+\gA\gB\right)}.
\end{equation}
This allows us to identify the teleportation fidelity possible for different choices of machine parameters. In particular, as we now show, there exists no choice of machine parameters that enables a larger singlet fraction than that obtained in the above example.

To this end, it is easily checked that in the limits $g\rightarrow 0$ and $g\rightarrow \infty$, one has a trivial $F\leq \frac{1}{2}$.  Next, we use the fact that the steady state is identical if all the parameters $ (g,\gA,\gB)$ are multiplied by a positive constant as discussed earlier in the paper. This allows us to fix $\gA=1$ without loss of generality. Then, we solve $\frac{\partial F}{\partial g}=0$ and obtain a single relevant solution, namely
\begin{equation}\label{gsol}
g=\frac{\sqrt{1+4\gB+10\gB^2+4\gB^3+\gB^4}-1-\gB^2}{4(1+\gB)}.
\end{equation}
Inserting this back into \eqref{fidneg}, we are left only to optimise over $\gB$, which is achieved again by considering the roots of the derivative. One finds $\gB=1$, which via \eqref{gsol} returns the optimal teleportation strategy presented earlier.

\subsection{Violation of the CHSH inequality is impossible}
Population inversion is, however, not a sufficiently strong resource to generate entanglement that can violate the CHSH inequality. To prove this, we show that the steady state \eqref{steady}  satisfies the condition \eqref{chshcond}. Thus, we define $S:=8\alpha^2+\left(2\Delta-1\right)^2$ and show that it can never exceed the value $1$. Differentiating this expression w.r.t $g$ one finds only two relevant roots, namely $g=0$ and 
\begin{equation}
g=\frac{|\gA^2-\gB^2|}{2\sqrt{\frac{\gA^3}{\gA}+\frac{\gB^3}{\gA}-4t^2+6\gA\gB}}.
\end{equation}
For the first root, one immediately obtains $S=1$. To investigate the second root, we  use again properties of our steady state \eqref{form} to fix $\gA=1$ without loss of generality. The task then simplifies into maximising the one-variable function
\begin{equation}
S=\frac{2\gB\left(1-4\gB+\gB^2\right)}{1-2\gB-2\gB^2-2\gB^3+\gB^4}.
\end{equation}
Calculating the roots of the derivative, one obtains $\gB\in\{-1,1,3-2\sqrt{2},3+2\sqrt{2}\}$. It is easily checked that each of these roots (as well as the limit $\gB\rightarrow \infty$) satisfy $S\leq 1$. Lastly, we check the limit $g\rightarrow \infty$, in which one obtains $S=\left(\frac{\gA-\gB}{\gA+\gB}\right)^4\leq 1$. Thus, it holds that $S\leq 1$ which via \eqref{chshcond} implies that the CHSH inequality is satisfied. This constitutes our second no-go theorem, stating that nonlocality can not be achieved neither with the simplest machine, nor with population inversion.

\subsection{Steerability is enhanced}
\begin{figure}[t!]
	\centering
	\includegraphics[width=\columnwidth]{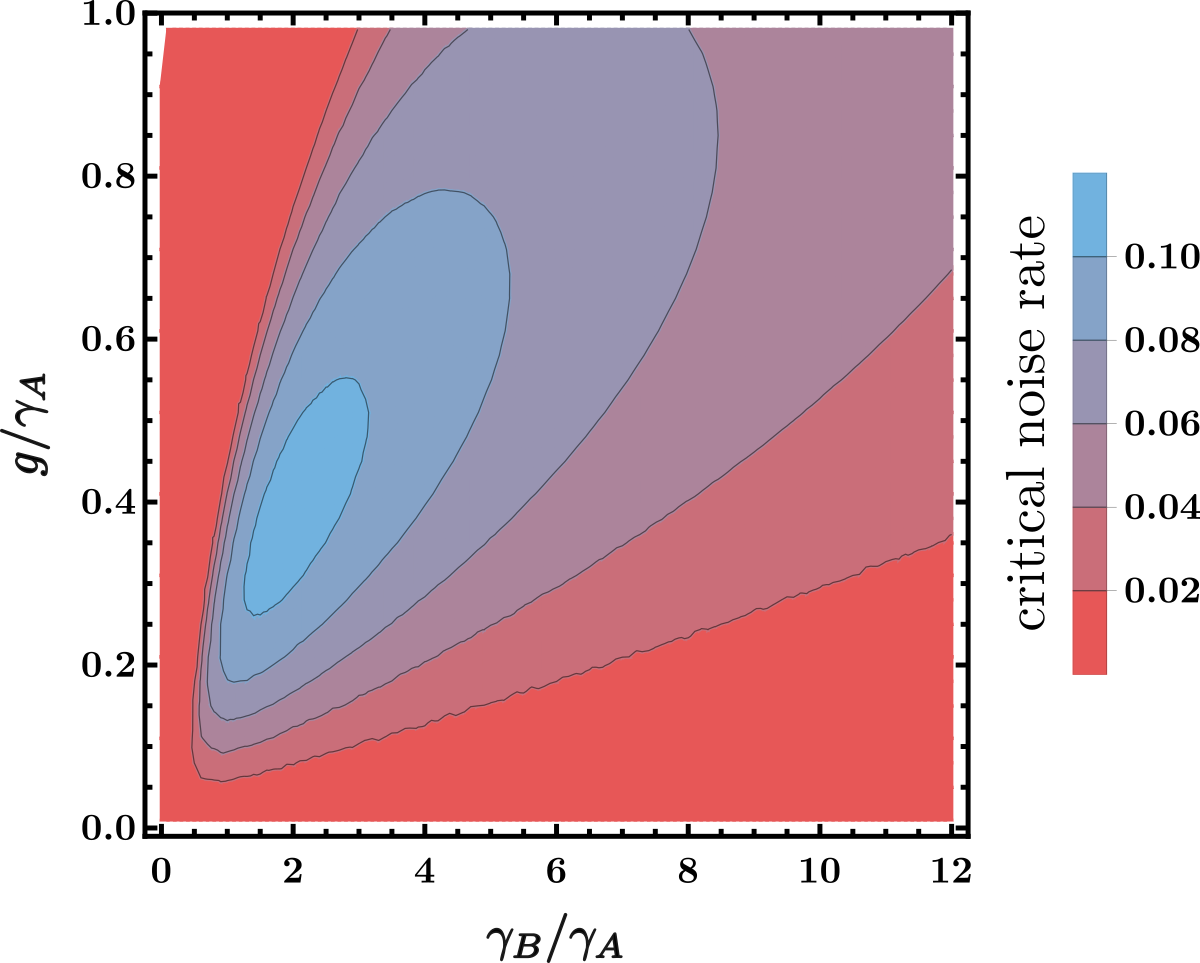}
	\caption{Strength of steering  vs.\ coupling rates in machine with population inversion, as measured by the critical isotropic noise rate $q$ in the steady state \eqref{steady} using the dodecahedral measurement configuration on qubit A.}\label{FigSteeringInvertedMain}
\end{figure}

Since we already found that the simplest machine enables steering, the same trivially follows for the present machine. In Figure~\ref{Figsteeringfermion}b we illustrate the (un)steerable region, in the space of the ratios $\frac{g}{\gA}$ and $\frac{\gB}{\gA}$ that characterise the steady state \eqref{steady}, using the method of Ref.~\cite{Nguyen2019}. The parameter region corresponding to a steerable steady state is found to be larger than in the previous cases. Moreover, as we show here, population inversion quantitatively enhances steerability. Concretely, steering can be demonstrated using the  dodecahedral measurement configuration (see section~\ref{secsteering}) at levels of isotropic noise that are  one order of magnitude higher than that tolerated in the previously considered machines. We have made a grid in the plane of $\left(\frac{g}{\gA}, \frac{\gB}{\gA}\right)$ and for each point evaluated the largest noise rate $q$ that is compatible with a local hidden state model given Alice's measurement strategy. To this end, we solve the semidefinite program \eqref{steeringsdp}, with the minor modification that $\max q$ is introduced as an objective function. In Figure~\ref{FigSteeringInvertedMain}, we illustrate the isotropic noise tolerance of steerability enabled by the machine. We find that for appropriate choices of machine parameters, steering can be achieved at a noise rate of  $q\approx 10.9\%$ (at $\frac{g}{\gA}=0.38$ and $\frac{\gB}{\gA}=1.9$).  It is also interesting to note that the state optimal for teleportation is not the most noise-tolerant in terms of steerability. The noise tolerance of that state under the dodecahedral strategy is $q\approx 8.1\%$.

\section{Nonclassicality in machines with heralding}\label{secHerald}
In this section, we consider supplementing the machine with local filters. This means that once the steady state is generated, each qubit may be subject to a measurement. If that measurement returns the desired outcome (i.e.~the state passes the filter), then we consider the post-measurement two-qubit state as final output state of the process. This  allows one to boost entanglement at the price of only probabilistically completing the procedure.

We may write the filters as the following measurements $\{F_\text{A}^\dagger F_\text{A},\openone-F_\text{A}^\dagger F_\text{A}\}$ and $\{F_\text{B}^\dagger F_\text{B},\openone-F_\text{B}^\dagger F_\text{B}\}$ on qubit A and qubit B respectively. We choose the Kraus operators to filter in the energy eigenbasis; $F_k=a_k\ketbra{0}{0}+b_k\ketbra{1}{1}$ for $k\in\{\A,\B\}$, for some coefficients $0\leq (a_\A,b_\A,a_\B,b_\B)\leq 1$. The heralded state becomes
\begin{equation}\label{filter}
\rho_\text{herald}=\frac{1}{p_\text{suc}}\left(F_\text{A}\otimes F_\text{B}\right) \rho_\text{steady}\left(F_\text{A}^\dagger\otimes F_\text{B}^\dagger\right),
\end{equation}
where $p_\text{suc}$ is the heralding efficiency given by
\begin{equation}
p_\text{suc}=\Tr\left(F_\text{A}^\dagger F_\text{A}\otimes F_\text{B}^\dagger F_\text{B}\rho_\text{steady}\right).
\end{equation}
The heralded state takes the form
	\begin{widetext}
	\begin{equation}
	\rho_\text{herald}=\frac{1}{M}
	\left(
	\begin{array}{cccc}
	4 a_1^2 a_2^2 g^2 \gB^2 & 0 & 0 & 0 \\
	0 & 4 a_1^2 b_2^2 g^2\gA \gB & -2 i a_1 a_2 b_1 b_2 g \gA \gB t & 0 \\
	0 & 2 i a_1 a_2 b_1 b_2 g \gA \gB t  & a_2^2 b_1^2 \gA \gB \left(4 g^2+t^2\right) & 0 \\
	0 & 0 & 0 & 4 b_1^2 b_2^2 g^2 \gA^2 \\
	\end{array}
	\right),
	\end{equation}
	\end{widetext}
where $M=t\left(4b_2^2 g^2 \gA+a_2^2\gB\left(4g^2+\gA t\right)\right)$.
Next, we  investigate the relationship between nonclassicality and this efficiency.

\subsection{Heralding in the simplest machine}
We now consider that $\rho_\text{steady}$ is the steady state obtained from the simplest machine, based on either bosonic or fermionic baths and interactions with and without  a charge term. Recall that although the simplest machine did enable steering, teleportation and nonlocality were out of reach. Therefore, we numerically explore the trade-off between the singlet fraction (for teleportation) and the CHSH parameter (for nonlocality) respectively, against the heralding efficiency $p_\text{suc}$. 

In both cases, we numerically find that in the asymptotic limit $p_\text{suc}\rightarrow 0$ the machine is able to generate a state that is essentially identical to the  maximally entangled state. This shows that in principle, the simplest autonomous machine supplemented with local filters enables optimal teleportation and a maximal violation of the CHSH inequality. However, to also account for a practically relevant scenario, it is important to investigate the trade-off between nonclassicality and $p_\text{suc}$.

\begin{figure}[t!]
	\centering
	\includegraphics[width=0.85\columnwidth]{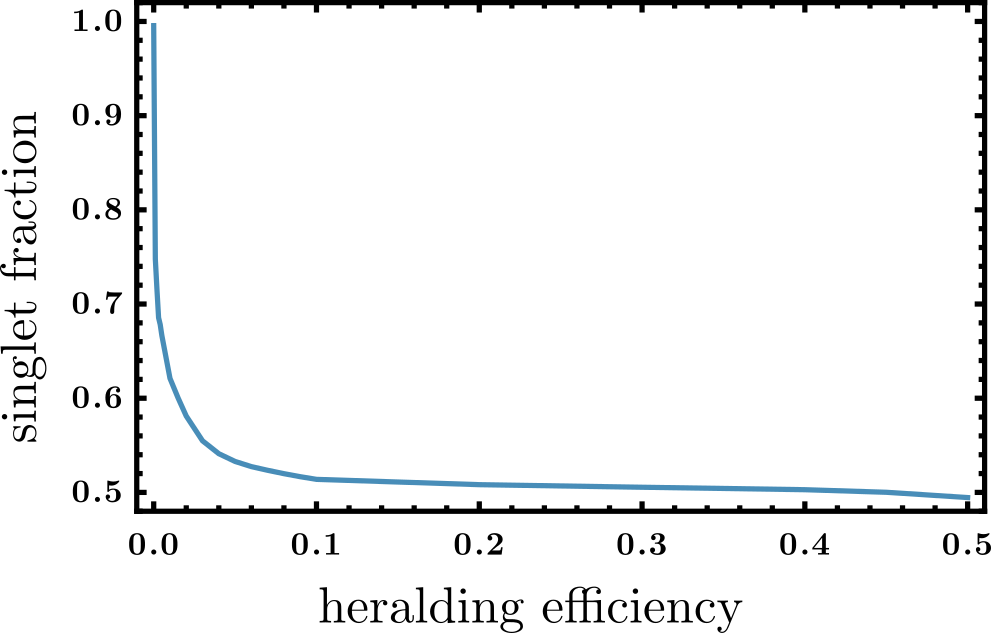}
	\caption{Usefulness for teleportation, as measured by the singlet fraction, vs.\ heralding efficiency for the autonomous machine with bosonic baths. The optimisation is performed in the limit $\TB\rightarrow 0$ and $u\rightarrow \infty$ over the machine parameters $(g,\gA,\gB,\TA)$ and the filter parameters $(a_\A,b_\A,a_\B,b_\B)$.}\label{FigHeraldedBoson}
\end{figure}

In the case of bosonic baths, we find that the robustness of the nonlocality is very small. Already at $p_\text{suc}=0.2\%$, we are no longer able to herald a state that can violate the CHSH inequality. Somewhat surprisingly, the situation for teleportation is very different. In Fig.~\ref{FigHeraldedBoson} we illustrate the numerical results for the singlet fraction versus the heralding efficiency. Although the the singlet fraction rapidly drops from the ideal value when $p_\text{suc}$ is perturbed away from the asymptotic limit, the rate of decrease also rapidly flattens around $p_\text{suc}\approx 10\%$. This allows us to detect a small but nontrivial singlet fraction up to $p_\text{suc}\approx 45\%$.

\begin{figure}[t!]
	\centering
	\includegraphics[width=\columnwidth]{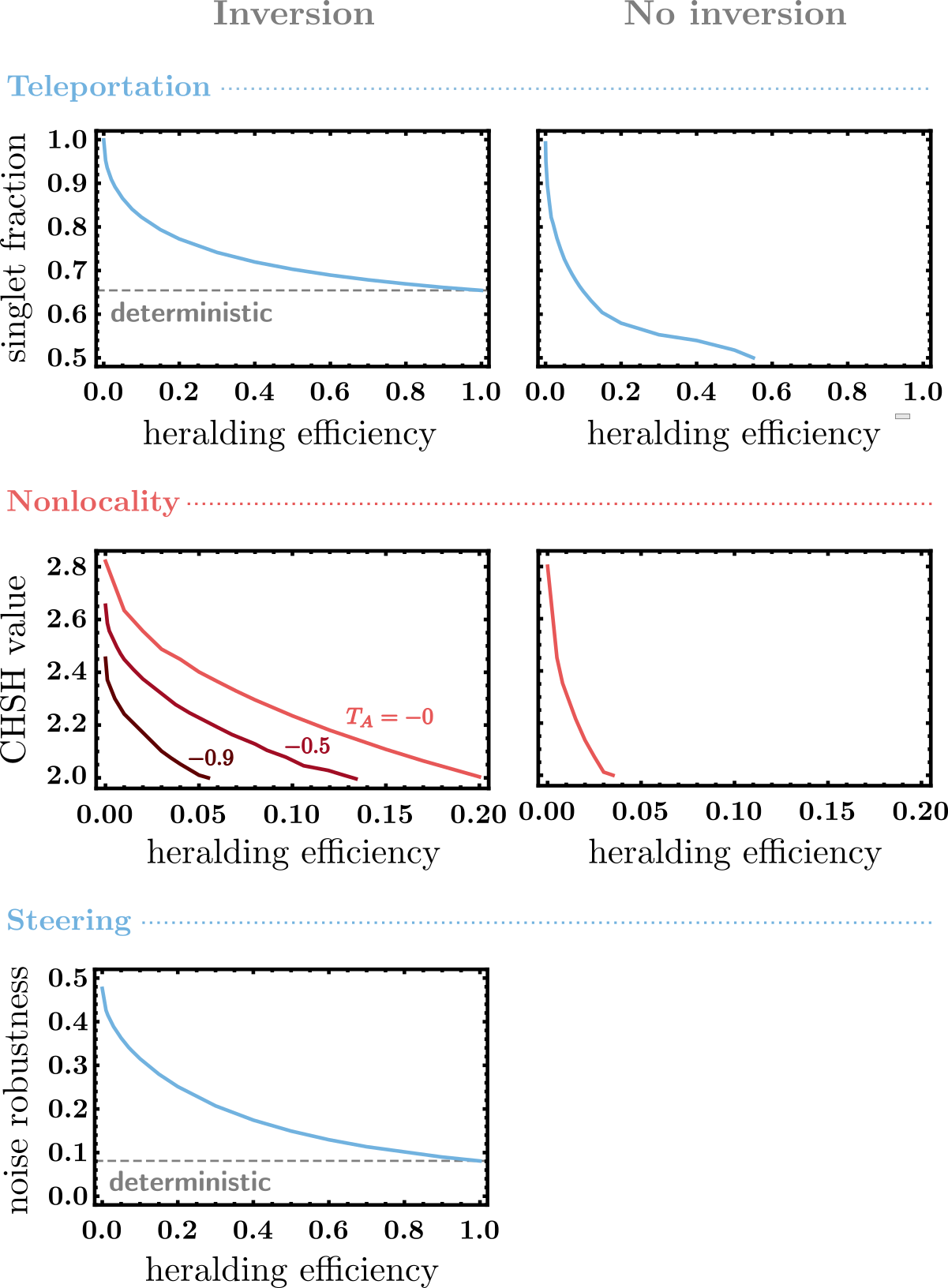}
	\caption{Comparison of operational nonclassicality for machines with fermionic baths, local filtering, and with (left column) and without (right column) population inversion. Numerical optimisation is performed in the limit $\TB\rightarrow 0$ and $u\rightarrow \infty$ over the machine parameters $(g,\gA,\gB,\TA)$ and the filter parameters $(a_\A,b_\A,a_\B,b_\B)$.  The optimisation of the singlet fraction (top row) is performed in the limit $u\rightarrow \infty$ with a filter only on qubit B. The CHSH parameter (middle row) is considered at $u=20$, for $\TA=0$ without inversion and for different negative temperatures $\TA$ with inversion corresponding to having $100\%$, $88\%$, and $75\%$ of the population in the excited state.	The optimal state for teleportation is used to evaluate the isotropic noise robustness (the parameter $q$) of steering for a dodecahedral measurement configuration (bottom row). }\label{FigHeraldedFermionColumb}
\end{figure}

In the case of charged fermions, we have considered the limit $\TB\rightarrow 0$ and $u\rightarrow \infty$. For a given value of $p_\text{suc}$, we have numerically searched over the machine parameters $(g,\gA,\gB,\TA)$ and the filter parameters  $(a_\A,b_\A,a_\B,b_\B)$ to maximise the singlet fraction and the CHSH parameter respectively. The results are displayed in Fig.~\ref{FigHeraldedFermionColumb} (left, top \& middle). The robustness of the singlet fraction to sizable efficiencies is more pronounced than in the case of bosons. We find a nontrivial singlet fraction up to  $p_\text{suc}\approx 55\%$ and a violation of the CHSH inequality up to $p_\text{suc}\approx 3.5\%$. This is again an improvement on the bosonic case, but in general the efficiencies for generating nonlocality or strongly nonclassical singlet fractions remain small.

\subsection{Heralding with population inversion}
The low efficiencies tolerated in the above scheme motivate an exploration of an even more powerful machine, which features both local filters and population inversion in the bath. Therefore, we consider the application of local filters to the steady state in \eqref{steady}. 

Before addressing the trade-off between nonclassicality and efficiency, let us present a simple argument for the fact that in the limit $p_\text{suc}\rightarrow 0$, the machine can generate a maximally entangled state. To this end, we leverage the fact that every pure entangled two-qubit state can be probabilistically converted into a maximally entangled state through a local filter. Given this knowledge, we need only to show that for any given degree of purity close to unit, the machine can produce an entangled steady state. To show this, we take $g\ll (\gA,\gB)$ and consider a series expansion of the purity in $g$,
\begin{equation}
\Tr(\rho_\text{steady}^2)=1-\frac{8g^2\left(\gA^2+\gB^2\right)}{\gA\gB t^2}+\mathcal{O}(g^3).
\end{equation}
Also, we consider the entanglement concurrence \cite{Wootters1998, Hill1997}, which for a state of the form \eqref{form} becomes $\mathcal{C}(\rho)=2\left(|\alpha|-\sqrt{a_1\left(1-a_1-a_2-a_3\right)}\right)$. A series expansion gives
\begin{equation}
\mathcal{C}(\rho_\text{steady})=\frac{4g}{t}-\frac{8g^2}{t^2}+\mathcal{O}(g^3).
\end{equation}
We see that when $g\rightarrow 0$, the concurrence approaches zero linearly and the purity approaches one quadratically. Any purity sufficiently close to one can be achieved with a non-vanishing concurrence. Consequently, this state can be filtered into a state correspondingly close to the maximally entangled state.


Now, we investigate steering, teleportation and nonlocality versus the heralding efficiency. We choose to only apply a filter to qubit B ($F_\A=\openone$). The reason is that we have found that also filtering also on qubit A only yields a marginal improvement. Therefore, a simpler setup, with a single local filter, may be preferable. For fixed $p_\text{suc}$, we have numerically optimised the singlet fraction over the machine parameters $(g,\gA,\gB)$ and the filter parameters $(a_\B,b_\B)$ based on the steady state \eqref{steady}. Then, we have taken  the optimal state found for teleportation and evaluated its white noise tolerance for steering via the semidefinite program \eqref{steeringsdp} using the dodecahedral measurement configuration. The results are displayed in Fig.~\ref{FigHeraldedFermionColumb} (right, top \& bottom). As expected, we find optimal teleportation and nearly optimal steering (the small suboptimality is due to the choice of the dodecahedral measurement configuration)  in the limit $p_\text{suc}\rightarrow 0$. The singlet fraction and the noise robustness of steering both decrease gradually until they reach the values reported in section~\ref{secnegtemp} for the deterministic setting ($p_\text{suc}=1$). 

We have similarly optimised the CHSH parameter numerically. In this case, we have considered values away from the idealised limits for the charge interaction and the temperature gradient. Specifically, we have fixed $u=20$ and considered three different choices of negative temperature $\TA$,  each of which corresponding to a degree of population inversion. The results are illustrated in Fig.~\ref{FigHeraldedFermionColumb} (right, middle). We find that for complete population inversion ($\TA\rightarrow 0^-$), the ability to violate the CHSH inequality is nearly one order of magnitude more robust than in the previous case without population inversion: nonlocality is sustained up to heralding efficiencies of $p_\text{suc}\approx 20\%$. This gradually decreases as the degree of population inversion is reduced.


\section{Conclusions}\label{secConc}
In this work, we have investigated the abilities and fundamental limitations in generating operationally useful entanglement from the steady state of minimal autonomous thermal machine. We found that even the simplest such machine is able to generate steady state entanglement that is strong enough for revealing steering, that the introduction of population inversion as an additional resource can boost entanglement to the extent that also teleportation becomes possible, and that the introduction of local filtering operations to herald entanglement from the steady state can enable Bell nonlocality at efficiencies up to about $20\%$.

A natural next step is to investigate operational nonclassicality and its limitations in autonomous entanglement engines based on either on more than two qubits or two higher-level systems. Also, taking the opposite point of view, it would be interesting to address to what extent one must introduce non-autonomous resources (e.g.~work input or time-dependent interactions) in order to generate entanglement that is strong enough to circumvent our no-go theorems on teleportation and CHSH violation in the two-qubit scenario. Here, we showed that local filtering is one possible avenue to such stronger entanglement.

Finally, our focus here has been exclusively on quantum thermal machines that perform a task that is inherently nonclassical, i.e.~entanglement generation. However, many quantum thermal machines are believed to outperform their classical counterparts at tasks that in themselves are classical (heating, cooling, extracting work etc.). Can these machines also manifest operationally meaningful nonclassicality?

\begin{acknowledgments}
AT was supported by the Swiss National Science Foundation through Early PostDoc Mobility fellowship P2GEP2 194800 and the Wenner-Gren Foundations. JBB  was supported by the Independent  Research Fund Denmark. FC is supported by the ERC Synergy grant HyperQ (Grant No. 856432), and GH acknowledges support from the Swiss national Science Foundation through the starting grant PRIMA PR00P2$\_$179748 and the NCCRs Quantum Science and Technology and SwissMAP.
\end{acknowledgments}
 
\bibliography{references_engine}

\appendix
\onecolumngrid


\section{Simplest machine with bosons}
\label{AppBosons}
The solution to the steady state Lindblad equation in the limit $\TB\rightarrow 0$ is given below on the form \eqref{form}. 
\begin{align}\label{bosonic}\notag
&a_1=\frac{4g^2\gA^2}{\left(\gA-\gB+t e^{\frac{1}{\TA}}\right)^2\left(4g^2+\gA\gB\coth\frac{1}{2\TA}\right)}\\\notag
& a_2=\gA\frac{\left(4g^2+\left(\gA-\gB\right)^2\right)\gB+te^{\frac{2}{\TA}}\left(4g^2+\gB t\right)-2e^{\frac{1}{\TA}}\left(-\gA^2\gB+\gB^3+2g^2 s\right)}{\left(4g^2\left(-1+e^{\frac{1}{\TA}}\right)+\gA\gB\left(1+e^{\frac{1}{\TA}}\right)\right)\left(\gA-\gB+te^{\frac{1}{\TA}}\right)^2}\\\notag
& a_3=4g^2\gA\frac{t+\gA\left(-1+e^{\frac{1}{\TA}}\right)^{-1}}{\left(4g^2\left(-1+e^{\frac{1}{\TA}}\right)+\gA\gB\left(1+e^{\frac{1}{\TA}}\right)\right)\left(\gB+\gA \coth \frac{1}{2\TA}\right)^2}\\
& \alpha=-\frac{2ig\gA\gB\left(-1+e^{\frac{1}{\TA}}\right)}{\left(4g^2\left(-1+e^{\frac{1}{\TA}}\right)+\gA\gB\left(1+e^{\frac{1}{\TA}}\right)\right)\left(\gA-\gB+te^{\frac{1}{\TA}}\right)},
\end{align}
where we have defined $t=\gA+\gB$ and $s=\gA+2\gB$. Observe that this state is invariant under the transformation $(g,\gA,\gB)\rightarrow c\times (g,\gA,\gB)$, for any $c>0$. Note also that the above is a special case of the general solution to the Lindblad for two interacting qubits coupled individually to their own bath, as derived in Ref.~\cite{Khandelwal2020} for fermions and bosons.

\subsection{Teleportation and CHSH violation is impossible}
We prove that $\Delta\leq \frac{1}{2}$ for all machine parameters. Some simplifications give
\begin{equation}
\Delta=\gA\frac{\left(8g^2+\left(\gA-\gB\right)^2\right)\gB+te^{\frac{2}{\TA}}\left(8g^2+\gB t\right)-2e^{\frac{1}{\TA}}\left(-\gA^2\gB+\gB^3+4g^2 s\right)}{\left(4g^2\left(-1+e^{\frac{1}{\TA}}\right)+\gA\gB\left(1+e^{\frac{1}{\TA}}\right)\right)\left(\gA-\gB+te^{\frac{1}{\TA}}\right)^2}.
\end{equation}
As previously discussed, we may fix, for instance, $\gA=1$. Then, differentiating w.r.t. $g$, one finds a single root at $g=0$, which corresponds to $\Delta=\left(1+e^{\frac{1}{\TA}}\right)^{-1}$ which is always smaller than $\frac{1}{2}$. We must also check the (unphysical, w.r.t. our master equation) limit $g\rightarrow \infty$, where one finds
\begin{equation}
\Delta=\frac{2e^{\frac{1}{\TA}}\left(1+\gB\right)-2\gB}{\left(1-\gB+e^{\frac{1}{\TA}}\left(1+\gB\right)\right)^2}.
\end{equation}
Differentiating w.r.t $\gB$, one finds a single root at $\gB=-1$ (unphysical), which corresponds to a maximum. Fixing this value eliminates $\TA$ and obtains  $\Delta=\frac{1}{2}$. Hence, for every choice of machine parameters $(g,\gA,\gB,\TA)$, we have $\Delta\leq \frac{1}{2}$ which implies that the singlet fraction and the CHSH value are both trivial.

\section{Simplest machine with charged fermions}\label{AppColoumb}

\subsection{Teleportation is impossible}
We work in the most relevant limit in which $\TB\rightarrow 0$ and $u\rightarrow \infty$ and prove that $F\leq \frac{1}{2}$ for any choice of $(g,\gA,\gB,\TA)$. To this end, we evaluate $F=\alpha+\frac{a_2+a_3}{2}$ for the steady state when transformed into the form \eqref{form2}. Some simplification gives 
\begin{equation}
F=\frac{\gA \left(8 g^2 \left(e^{\frac{1}{\TA}}+1\right)+4 g \left(e^{\frac{1}{\TA}}+1\right) \gB+\gB \left(te^{\frac{1}{\TA}}+\gB\right)\right)}{2
	\left(e^{\frac{1}{\TA}}+1\right) \left(4 g^2 \left(\left(e^{\frac{1}{\TA}}+1\right) \gB+\left(e^{\frac{1}{\TA}}+2\right) \gA\right)+\gB \gA \left(te^{\frac{1}{\TA}}+\gB\right)\right)}.
\end{equation}
For simplicity, one may define $x=\frac{1}{\TA}$. Evaluating the derivate w.r.t $x$, one finds that it is non-positive. Thus, the above right-hand-side is strictly increasing with $\TA$. Thus, we take the limit $\TA\rightarrow \infty$ and obtain
\begin{equation}\label{Fstep}
F=\frac{\gA \left(16 g^2+8 g \gB+\gB s\right)}{4 \left(4 g^2 (2t+ \gA)+\gA \gB s\right)}.
\end{equation} 
We take the derivative w.r.t. $g$ and find its roots. There is only one positive root, given by
\begin{equation}
g=\frac{\gA^2-4 \gB^2+\sqrt{s \left(\gA^3+46 \gA^2 \gB+28 \gA
		\gB^2+8 \gB^3\right)}}{8 (2t+\gA)}.
\end{equation} 
Inserting this in \eqref{Fstep} we obtain an expression only in terms of $(\gA,\gB)$. Once again, we can w.l.g. fix $\gA=1$. This gives
\begin{equation}
F=\frac{7+4 \gB (\gB+4)+\sqrt{(2 \gB+1) (2 \gB (2 \gB (2 \gB+7)+23)+1)}}{8 (4\gB (\gB+2)+3)}.
\end{equation}
Taking the derivative, its root can be evaluated analytically but is too cumbersome to state here. Its decimal approximation is $\gB\approx 0.300$ and gives $F\leq 0.3788$, which is below the critical threshold of $F=\frac{1}{2}$.

\subsection{CHSH inequality violation is impossible}
We work in the most relevant limit in which $\TB\rightarrow 0$ and $u\rightarrow \infty$ and prove that the CHSH inequality is satisfied for any choice of $(g,\gA,\gB,\TA)$. To this end, we use the previous result that the machine does not enable a non-trivial singlet fraction, i.e.~$F\leq \frac{1}{2}$, which via \eqref{telecond2} implies $a_2+a_3\leq 1-2\alpha$. Inserted into our condition \eqref{chshcond} for satisfying the  CHSH inequality, it reduces to  
\begin{equation}\label{chshlim}
8\alpha^2+\left(1-4\alpha\right)^2\leq 1,
\end{equation}
which is satisfied when $\alpha\leq \frac{1}{3}$.

In order to show that the machine always respects this condition, we prove a bound on $\alpha$ for all steady states. In the relevant limit ($\TB\rightarrow 0$ and $u\rightarrow \infty$), the magnitude of the off-diagonal element is
\begin{equation}
\alpha=\frac{2 \gA \gB g}{4 g^2 \left(\gA \left(e^{\frac{1}{\TA}}+2\right)+\gB
	\left(e^{\frac{1}{\TA}}+1\right)\right)+\gA \gB\left(\gB+te^{\frac{1}{\TA}} \right)}.
\end{equation}
Defining $x=\frac{1}{\TA}$ and calculating the derivative w.r.t. $x$, one finds that it is non-positive. Thus, to maximise $\alpha$, we take the limit $\TA\rightarrow \infty$. This gives
\begin{equation}
\alpha=\frac{2g\gA\gB}{\gA\gB s+4g^2\left(3\gA+2\gB\right)}.
\end{equation}
Differentiating w.r.t. $g$, one obtains the single positive root
\begin{equation}
g=\frac{1}{2}\sqrt{\frac{\gA\gB s}{3\gA+2\gB}},
\end{equation}
which leads to
\begin{equation}
\alpha=\frac{1}{2}\sqrt{\frac{\gA\gB}{3\gA^2+8\gA\gB+4\gB^2}}.
\end{equation}
Differentiating w.r.t. $\gA$, the single positive root is $\gA=\frac{2}{\sqrt{3}}\gB$. This leads to the following maximal value for the magnitude of the coherence term:
\begin{equation}
\alpha=\frac{\sqrt{2-\sqrt{3}}}{4}\approx 0.129.
\end{equation}
Thus, the condition \eqref{chshlim} is always satisfied.

\end{document}